\documentclass[12pt]{iopart}
\usepackage{iopams}
\usepackage{epsfig}
\begin{document}

\title[Interaction of matter-wave gap solitons in optical lattices]
{Interaction of matter-wave gap solitons in optical lattices}

\author{Beata~J.~D{\c a}browska, Elena~A.~Ostrovskaya and~Yuri~S.~Kivshar\footnote[3]{
Corresponding author. E-mail address: ysk124@rsphysse.anu.edu.au (Yuri Kivshar)}}

\address{Nonlinear Physics Centre and Australian Centre for Quantum-Atom Optics,
Research School of Physical Sciences and Engineering, Canberra ACT 0200, Australia}

\begin{abstract}
We study mobility and interaction of gap solitons in a
Bose-Einstein condensate (BEC) confined by an optical lattice
potential. Such localized wavepackets can exist only in the gaps
of the matter-wave band-gap spectrum and their interaction
properties are shown to serve as a measure of discreteness imposed
onto a BEC by the lattice potential. We show that inelastic
collisions of two weakly localized near-the-band-edge gap solitons
provide simple and effective means for generating strongly
localized in-gap solitons through soliton fusion.
\end{abstract}

\submitto{\JOB}


\maketitle

\section{Introduction}
\label{intro}

Periodic potentials created by optical lattices are by now
recognized as a powerful tool for controlling and manipulating
nonlinear matter waves through dynamical diffraction management
\cite{marcus_dm,inguscio_dm}. One of the important implications of
such management is creation of nonlinearly localized matter-wave
packets, the so-called {\em bright atomic solitons}, in a~Bose-Einstein 
condensate with a positive scattering length, i.e.
with {\em repulsive} inter-atomic interaction. Such a
counter-intuitive effect can occur as wave localization in the
gaps of the band-gap spectrum due to the Bragg scattering of
coherent matter waves by the periodic potential of an optical
lattice~\cite{Zob_et99}. Since the repulsive atomic interactions
prohibit a collapse of the bright wavepackets and the solitonic
nature of their evolution ensures fixed spatial phase variations
due to mean-field effects, the atomic gap solitons may represent
an attractive high-density source for atomic
interferometry~\cite{gap_control}.

The possibility of nonlinear localization in a~repulsive BEC was
established theoretically~\cite{Zob_et99,Kon_Sal02,lenaPRA,efrem},
being recently confirmed experimentally in the case of
one-dimensional optical lattices~\cite{gap_exp}. The experimental
challenge of the gap soliton generation and manipulation is
two-fold. First, the BEC wavepacket, initially loaded into a
ground (Bloch) state of the optical lattice potential, should be
accelerated to the edge of the first Brillouin zone. The regime of
the wavepacket preparation and the dynamical transition to the
band edge dramatically affects the outcome of the
experiment~\cite{nl_dynamics}, and in the best case scenario only
weakly localized low-atom-number gap solitons near the bottom edge
of the spectral gap can be generated using the current
experimental techniques~\cite{gap_exp}. The depth of the gap,
where the gap solitons contain large atom numbers and are well
localized~\cite{lenaPRA} is yet to be accessed. Secondly, the
possibility of the gap soliton manipulation, which is essential
for interferometric applications, can be limited by impaired
mobility of the gap solitons due to effects of lattice
discreteness~\cite{lewenstein}.

In this paper, we examine the problem of mobility and interactions
of gap solitons in a one-dimensional optical lattice
without an additional harmonic confinement, within the framework
of the continuous Gross-Pitaevskii model with a periodic
potential. We show that, as the chemical potential of a BEC scans
the spectral gap, the gap solitons display a variety of scattering
properties ranging from those of ``ideal'' solitons without a
periodic potential to those of nonlinear localized excitations of
discrete nonlinear lattices~\cite{malomed}. In addition, we
suggest that the inelastic collisions of near-the-band-edge gap
solitons can provide reliable means for generating
highly-localized high-density immobile atomic solitons in the
depth of the spectral gap.

\section{Model}
\label{model}

We~consider the~dynamics of~a cigar-shape~BEC cloud
in~the~presence of~a one-dimensional ~optical lattice.  In
dimensionless units the~Gross-Pitaevskii (GP)~equation governing
the evolution of the condensate wavefunction can~be~written as:
\begin{equation}
    \label{GPE1D}
         i {\partial \psi\over \partial t}=
         \left( -{1\over 2}{\partial^2\over\partial x^2}+ V(x) +\sigma |\psi|^2\right)\psi,
\end{equation}
where the~optical lattice potential $V(x)=V_0{\rm sin}^2(\pi x/d)$
is~uniform and characterized by~its~depth $V_0$ measured in units
of lattice recoil energy $E_r=\hbar^2 \pi^2/(2md^2)$, and~lattice
period $d$. The~condensate wave function is~normalized to: $
\int_{-\infty}^{+\infty}dx|\psi(x,t)|^2=Ng_{1D}$, where $N$
is~the~number of~atoms, $g_{1D}=2|a_s|/a_0$, $a_s$~is~the~s-wave
scattering length, and $a_0$~the~characteristic harmonic oscillator
length in the direction of tight confinement. For~$^{87}$Rb atoms
with $m=1.44\times10^{-25}$kg, $a_s=5.3$nm and assuming a transverse
frequency $\omega_{\perp}\sim2\pi\times10^{2}$Hz, gives
$g_{1D}\sim 10^{-2}$. The~coefficient $\sigma=sgn(a_s)=\pm1$
characterizes the~type of~the~atomic interactions. In what follows
we consider repulsive interaction, i.e.  $\sigma=1$, and set
$d=1$, which corresponds to $\approx 1.1$ $\mu m$ lattice spacing.

\begin{figure}[H]
\centering \epsfig{figure=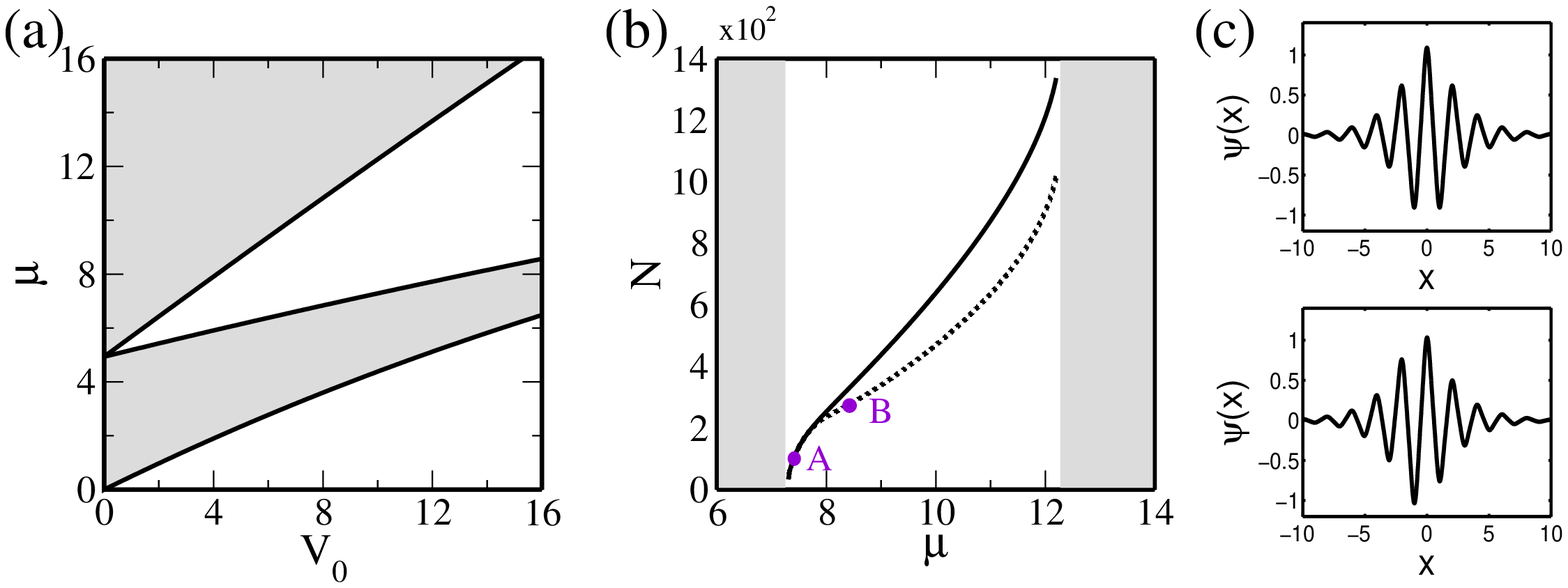,scale=0.65} \caption{\emph{(a)
Band-gap structure of the matter-wave spectrum $\mu(V_0)$. (b)
Families of the on-site (OS, dotted) and inter-site (IS, solid) gap solitons for
$V_0=10.0$, within the first gap. Point A (B) corresponds to the
location of an initial (final) state in~\fref{figure_5}. (c)
Spatial profiles of the OS (top) and IS (bottom) gap solitons at
$\mu=7.5$.}} \label{figure_1}
\end{figure}

Stationary states of~a~condensate are~described by~solutions
of~Eq.(\ref{GPE1D}) of~the~form: $\psi(x;t)=\phi(x)\exp(-i\mu t)$,
where the~ steady-state wave function $\phi(x)$ obeys the~{\it
time-independent} GP~equation
\begin{equation}
\label{TIGPE1D}
\left( -{1\over 2}{d^2 \over dx^2}+ V(x) +\sigma
|\phi|^2\right)\phi=\mu\phi.
\end{equation}
In~the~case of~a~noninteracting condensate ($\sigma=0$) 
stationary solutions of~equation~\eref{TIGPE1D} are Bloch waves,
$\phi(x)\sim b(x,k){\rm exp}(ikx)$, where $b(x)$ have periodicity
of the lattice and $k$ is the quasi-momentum. The~linear spectrum
$\mu(k)$ of the Bloch states has a characteristic band-gap
structure, with the lowest two bands shown in~\fref{figure_1}(a)
in the parameter plane~($V_0$, $\mu$). The shaded areas
correspond to~the regions where oscillating solutions
of~equation~\eref{TIGPE1D} (Bloch waves) exist. The clear areas
correspond to the spectral gaps which appear due to the Bragg
scattering of the matter waves in a periodic structure, being 
the~forbidden domains for~matter wave propagation. The gap below 
the~first band is~the trivial semi-infinite "total internal
reflection" gap. The first finite gap is where the formation 
of~gap solitons in repulsive BEC occurs~\cite{lenaPRA,efrem}. The
condition $\mu=V_0$ roughly delineates between the regimes 
of~tight-binding ($\mu<V_0$) and superfluid ($\mu>V_0$) behavior. In
what follows, we investigate the dynamics of gap solitons 
in~the~lattice of height $V_0=10$ which places the first gap into a
relatively tight-binding regime.

The nonlinear localization of a~BEC with positive scattering length
($\sigma=+1$) in the first spectral gap has been well studied both
theoretically~\cite{lenaPRA,efrem} and
experimentally~\cite{gap_exp}. The stationary localized solutions
of equation~\eref{TIGPE1D} - gap solitons - can be found at any value
of $\mu$ inside the spectral gap. The number of atoms contained in
a localized state is small near the lowest gap edge and  raises as
the chemical potential moves inside the gap~\cite{lenaPRA}. The
lowest-order families of~{\it on-site} (OS) and~{\it inter-site}
(IS) gap solitons, centered on the lattice minimum or maximum,
respectively, are shown in~\fref{figure_1}(b). The characteristic
"staggered" spatial structure of both OS and IS solitons can be
seen in~\fref{figure_1}(c).

\section{Mobility of gap solitons}
\label{moblility}

In order to study interactions of gap solitons, one has to
generate mobile localized states. It has been shown, however, that
the~mobility of~a matter wave soliton in an optical lattice
is~restricted by~the existence of the so-called
Pei\-erls~Na\-bar\-ro (PN) potential
bar\-rier~\cite{lewenstein,malomed}. The concept of the PN barrier
originates from the theory of dislocations and later studies of
mobility of localized nonlinear excitations in discrete 
lattices~\cite{KivsharCampbell,braun}. The PN barrier is introduced as
the height of the effective periodic potential generated by the
lattice discreteness, and it defines the minimum energy required
to move the centre of mass of a localized wavepacket by one
lattice site~\cite{KivsharCampbell}. Extending this definition to
the continuous lattice model, one can define the PN barrier as the
energy difference between the on-site and inter-site localized
stationary states~\cite{lewenstein}. Since the OS and IS states
can be mapped to represent a moving mode at different times, the
PN barrier can be calculated as the~smallest amount of energy
that~a~gap soliton needs to~gain in~order to~start moving along
the~lattice. Even in the tight-binding regime, the amount of the
energy needed to initiate the soliton motion is not in agreement
with the estimate of the PN barrier height derived from the
corresponding discrete nonlinear Schr\"o\-din\-ger model
\cite{lewenstein}, therefore the use of the continuous model in
the study of gap soliton motion and interactions is well
justified.

\begin{figure}[H]
\centering \epsfig{figure=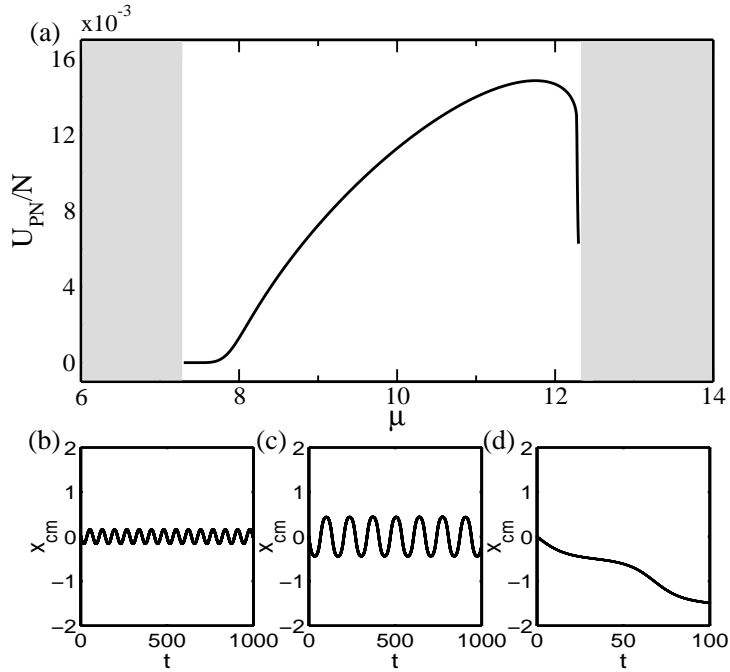,scale=0.45} \caption{\emph{Top:
(a) PN potential for gap solitons presented in~\fref{figure_1}(b)
as a function of $\mu$ within the first gap. Bottom: Examples of
soliton centres of mass dynamics for $\mu=7.7$ with the PN threshold momentum
$k_{PN}=1.20\times 10^{-2}$ and the initial momentum (b) below the
PN threshold, $k_0=6.01\times10^{-3}$, (c) near the threshold
$k_0=1.20\times10^{-2}$, and (d) above the threshold
$k_0=1.24\times10^{-2}$.}} \label{figure_2}
\end{figure}

\begin{figure}[H]
\centering \epsfig{figure=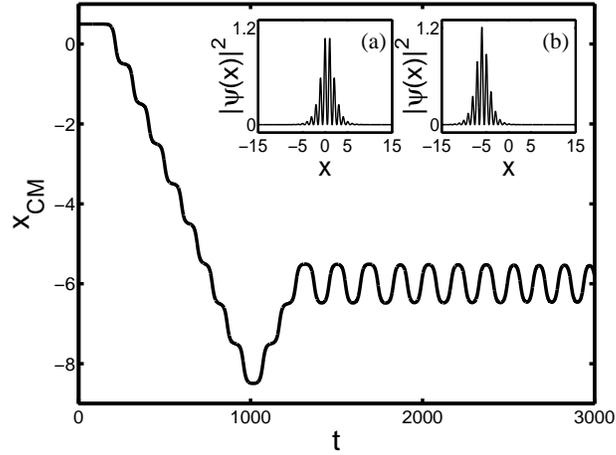,scale=0.8}
\caption{\emph{Evolution of the centre of mass of the initially
stationary IS gap soliton at $\mu=7.7$. Insets: (a) Initial
spatial profile, and (b) spatial profile at the time $t=3\times
10^3$, showing the charactersitic structure of the OS state.}}
\label{figure_IS}
\end{figure}

A soliton solution of~equation~\eref{TIGPE1D} is~characterized
by~its~dynamical invariants: number of~atoms and~energy.
The~soliton energy can be calculated as:
\begin{equation}
    \label{E}
         E=\int dx \left[-{1\over 2}\phi^{*}{d^2 \over dx^2}  \phi + V(x)|\phi|^2
         +{1\over2}\sigma |\phi|^4 \right].
\end{equation}
The on-site (OS) gap soliton is a localized state with the lowest
energy, and therefore it is this state that is eventually
generated from an arbitrary initial wavepacket. The PN potential
height associated with this state is given by
$U_{PN}=E_{IS}-E_{OS}$. The~threshold value of the momentum (per
atom) that a soliton must have to overcome the barrier is defined
as $k_{PN}=(2 U_{PN}/ N)^{1/2}$. Such a momentum can be~applied
experimentally either by~ultrashort laser pulses or a~tilt
of~the~optical lattice potential imposed by a gravitational field
or by a slope of a harmonic trap. The PN potential per atom
calculated for~the~family of~OS gap solitons within the first gap
is shown in~\fref{figure_2}(a). In contrast, the inter-site state
corresponds to a maximum of the PN potential and is {\em unstable
against small perturbations} of its center-of-mass position. Such
perturbation will lead to conversion of the IS soliton into an OS
soliton, as shown in \fref{figure_IS}.

To investigate the dynamics of the OS gap soliton with a finite
velocity, we solve the evolution equation~\eref{GPE1D}
numerically. The initial condition is given by 
$\psi(x;t=0)=\phi_0(x)\exp{(-i k_0 x)}$, where $\phi_0$ is a stationary gap
soliton wavefunction, and $k_0$ is an~initial momentum per atom.
The values of $k_0$ were chosen below, above, or at~the~threshold
value $k_{PN}$. The results of the calculations are shown in
\fref{figure_2}(b,c,d). Due to the large negative group-velocity
dispersion~\cite{drago}, solitons near the~edge of the~lower band
exhibit {\em anomalous steering}, i.e. initially the direction of
motion is opposite to the momentum kick received. Oscillating
motion within a~single lattice site has been~observed
for~an~initial $k_0$ values below and at~the~threshold
[see~\fref{figure_2}(b,c)]. Above the~threshold kinetic energy (which
is still an order of magnitude lower than an individual lattice
well depth) a~soliton undergoes free motion across the lattice, as
shown in~\fref{figure_2}(d).  During this~motion, an~atomic
soliton evolves through many states centered on~different points
of~the~optical lattice. Periodic revival of~the states closely
resembling~the~stationary states~of the IS and OS families
[\fref{figure_1}(c)] can be observed, however due to a loss
of~atoms neither number of~atoms contained in a~soliton nor $\mu$ 
are conserved during the evolution.
The loss of atoms from the moving localized state increases
with~growing $k_0$.

The~mobility properties of~gap~solitons vary depending on their
chemical potential. Close to the lowest gap edge ($\mu<7.5$ in our
case), gap solitons are well described by the effective nonlinear
Schr\"odinger equation (NLS) for the slowly varying envelope of a
Bloch state~\cite{Steel}. In this regime, solitons are expected to
behave like lattice-free envelope solitons of the integrable NLS
model. Indeed, they exhibit free motion for arbitrary small
initial momentum due to~very small  energy difference between
the~OS and IS stationary solutions of~equation~\eref{TIGPE1D}. These
states are weakly localized and~contain only about $100$~atoms.
Deeper in the gap ($7.5<\mu<8.2$), the mobility properties of
solitons are similar to those of localized solutions of the
discrete NLS equation, and can be~very well described
by~the~concept of~the~PN barrier. In the regime where the solitons
are strongly localized in the vicinity of a single lattice well
($\mu>8.2$), their dynamics is dominated by oscillations of
the~centre of~mass within a single lattice site for moderate
initial perturbations. The soliton motion with a large initial
velocity is accompanied by significant loss of~atoms from the
localized structure.

\section{Binary collisions of gap solitons}
\label{collisions}

Depending on the value of the chemical potential and degree of
localization, gap solitons display a~variety of~scattering
properties. In particular, we study binary collisions of in-phase
solitons numerically for a~range of~values of~the~chemical
potentials near the bottom edge of the gap, which is the
experimentally accessible region of soliton generation
\cite{gap_exp}. We keep the initial separation of soliton centers
the same in all cases and chose the same initial velocity above
the~PN threshold. Similar interaction conditions can be realized
by imposing an additional harmonic confinement onto the system.

\begin{figure}[H]
\centering \epsfig{figure=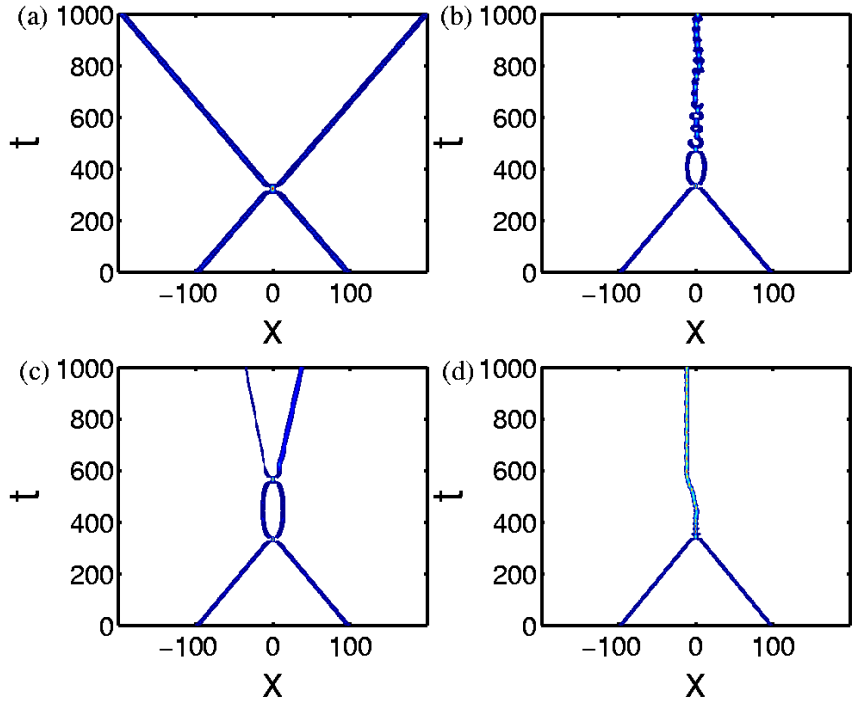,scale=0.45}
\caption{\emph{Examples of soliton interactions in the
nonintegrable NLS regime for different values of the chemical
potential (a) $\mu=7.35$; (b) $\mu=7.405$; (c) $\mu=7.41$; (d)
$\mu=7.45$ and  $k_0=0.1$.}} \label{figure_3}
\end{figure}

Three regimes of~soliton collisions have been identified in our
simulations. The first one is~the~regime where in-phase solitons
($\Delta\phi=0$) interact elastically, i.e. without momentum and
energy exchange [see \fref{figure_3} (a)]. In~this~narrow
parameter domain ($7.3<\mu<7.35$) the~dynamics of~the~gap~solitons
resembles the~dynamics of "ideal" solitons of~the~completely
integrable NLS equation without the lattice potential. The total
momentum of the system is defined as:
\begin{equation}
    \label{P}
    P=i\int dx \left({d\psi^*\over dx}\psi-\psi^*{d\psi\over dx}\right),
\end{equation}
and its zero value before and after collision indicates that the
soliton collisions are not affected by the lattice periodicity.

Deeper in~the~gap the~solitons start to experience stronger
localization and their dynamics become inelastic. Significant
radiation usually accompanies the collisions, and the~total
momentum of the system is no longer conserved. This is the typical
signature of a discrete system~\cite{malomed}, and a consequence
of nonintegrability introduced by a periodic potential. The~effect
of~small perturbations (numerical noise which emulates noise
in~the~real physical system) accounts for the~spontaneous
symmetry breaking (SB) and~generation of finite momentum during
the~collision process~\cite{malomed}. In~this~``weakly discrete''
regime of the soliton dynamics, the solitons can exhibit multiple
collisions~\cite{malomed}. In~the~multiple collisions solitons
"bounce" from each other leading to~either final separation
or~fusion, see~\fref{figure_3}(b,c). For $\mu\ge7.43$ the~solitons
merge into a~single localized state, as seen in
\fref{figure_3}(d), which due to~SB~moves in~an~unpredictable
direction across the~lattice and exhibits large amplitude
oscillations. The~appearance of~SB, multiple collisions
and~the~total momentum generation has~been~observed for a~range
of~the~chemical potential $7.35<\mu<7.5$.

\begin{figure}[H]
\centering \epsfig{figure=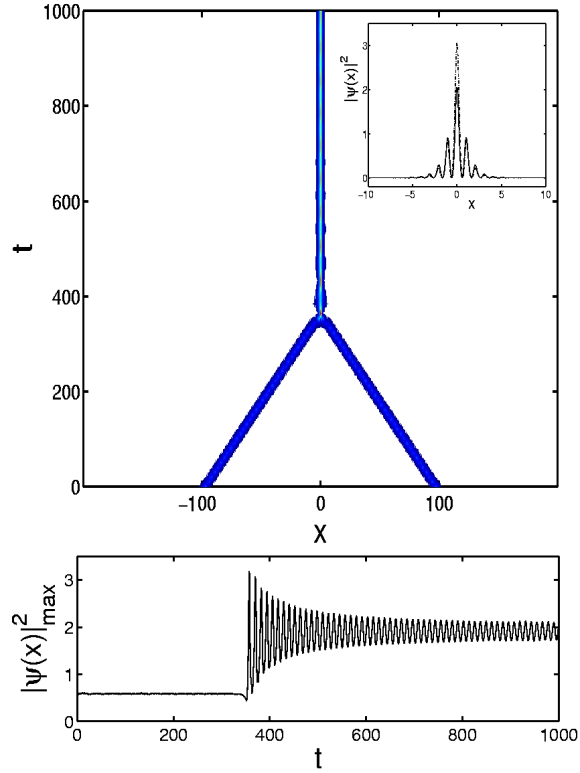,scale=0.35}
\caption{\emph{Top: Typical scenario of soliton fusion with
zero final velocity in the discrete NLS regime ($\mu=7.5$) 
Initial momentum of the soliton pair $|k_0|=0.1$.; 
Inset: Solid -
typical spatial profile of the post-collision trapped state at
$t=10^3$, dashed - exact OS solution corresponding to point B
in~\fref{figure_1}(b), $\mu=8.36$. Below: Evolution of the peak
density of the trapped state showing transition to a trapped
highly-localized in-gap state corresponding to the marked point B
in~\fref{figure_1}(b).}} \label{figure_4}
\end{figure}

\begin{figure}[H]
\centering \epsfig{figure=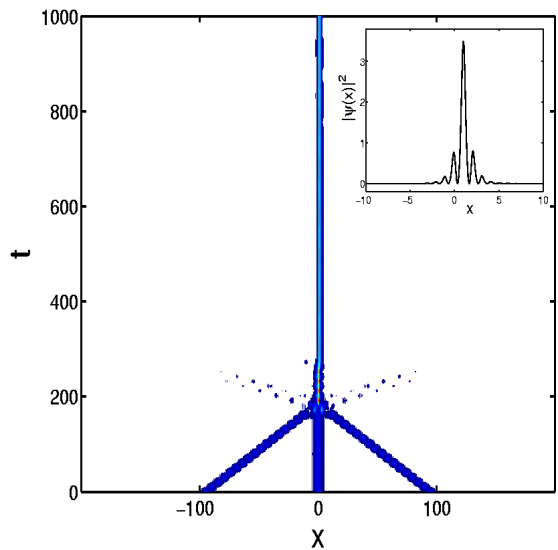,scale=0.35}
\caption{\emph{Typical scenario of triple-soliton fusion in the
discrete NLS regime ($\mu=7.5$), with the central immobile soliton $\pi$-out of phase with the
others. Initial momentum of the off-centre soliton pair $|k_0|=0.2$.; 
Inset: typical spatial profile of the
post-collision trapped state at $t=10^3$.}} \label{figure_5}
\end{figure}

For our choice of the optical lattice depth, the solitons
relatively close to the lower gap edge (e.g.~for~$\mu=7.5$)
already exhibit strongly inelastic scattering behavior
characteristic of a discrete system. For~the values of the~chemical
potential $\mu \geq 7.5$ the~binary collisions lead to~fusion of
two solitons into a single localized state. Unlike the outcome of
fusion after multiple collisions, the resulting solitons move with
nearly zero velocity and damped amplitude oscillations, as can be
seen in \fref{figure_4}. Location of~the~initial states
(containing around 170 atoms each) on the branches of stationary
solutions, is~indicated by~point A in~\fref{figure_1}(b). After
collision, an~immobile highly-localized soliton containing a
larger number of atoms is generated. From the $N$~vs.~$\mu$
dependence [see~\fref{figure_1}(b)] it can be inferred that
the~generated soliton, consisting of approximately 270 atoms, lies
deeper in the~gap than the~initial states [see point B
in~\fref{figure_1}(b)]. Therefore, inelastic collision of weakly
localized near-the-gap-edge solitons can, in principle, be employed
for~a~formation of fundamental atomic gap solitons with a larger
number of atoms in~the~depth of~the~spectral gap, i.e. in the
region of parameters that is currently inaccessible
experimentally. This technique can be useful for any number of
colliding solitons near the band edge. In particular, the case of
triple-soliton collisions is illustrated in \fref{figure_5}, were
the phase of the central soliton was inverted by $\pi$ to achieve
soliton merging.

\begin{figure}[H]
\centering \epsfig{figure=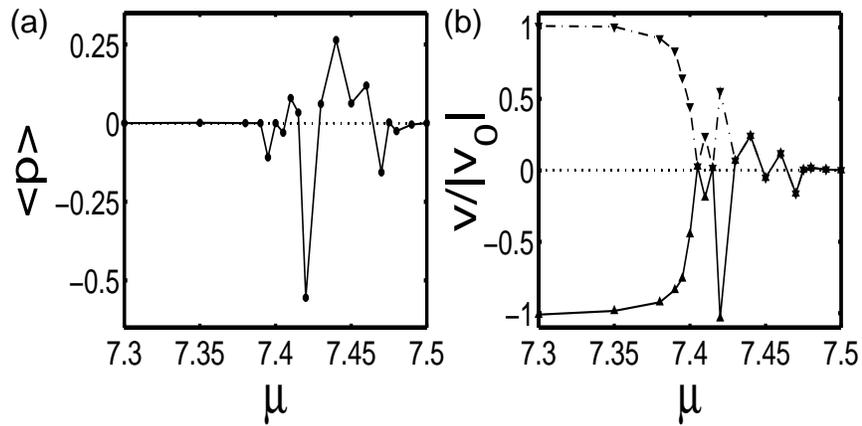,scale=0.45}
\caption{\emph{Post-collision (at $t=10^3$) (a) time-averaged
momentum of gap solitons as a function of $\mu$ within the first
gap and (b) relative velocity of the soliton pairs. Behavior
characteristic for $k_0=0.1$.}} \label{figure_6}
\end{figure}

For solitons located even deeper in the gap, the binary collisions
inevitably result in their fusion, and the emerging localized
states remain~confined within a~single lattice well, although
their~centre of~mass can oscillate around the~potential minimum.
The amplitude of~the~centre of~mass oscillations increases with~$\mu$,
however the~value of the total momentum averaged over the~time 
$\Delta t$ at $\sim 20$ oscillations:
$\langle p \rangle={1\over \Delta t}\int_{t_0}^{t_0+\Delta t} p dt$,
remains~close to zero. The values of the post-collision velocity
of the localized state (relative to the initial velocity $v_0$)
and the average momentum vs. chemical potential of the
colliding solitons in the vicinity of the gap edge are plotted 
in~\fref{figure_6}. It can be clearly seen that the initial stage of
quasi-elastic collisions ($\langle p \rangle \approx 0$, $v=v_0$)
and final stage of the soliton fusion ($\langle p \rangle \approx
0$, $v\approx 0$) is separated by the symmetry breaking stage,
where the outcome of collisions is not predictable. We note that
the symmetry breaking outcome of soliton collisions is also
characteristic of binary collisions of solitons with non-zero
phase difference, as in the case of a~localized state in the
discrete NLS model~\cite{malomed}.

\section{Conclusions}
\label{conclusions}

We have studied numerically both mobilility and interaction of the
matter-wave gap solitons in an experimentally accessible regime,
i.e. near the lowest edge of the spectral gap of an optical
lattice. For the collision velocities above the Peierls-Nabarro
mobility threshold, the gap solitons demonstrate scattering
dynamics scenarios ranging from quasi-elastic collisions to 
soliton fusion. By employing the inelastic collision of two or
three gap solitons in configurations that can be realized
experimentally, for instance, in a combination of an~optical lattice
and a harmonic trap, we showed that the soliton fusion can be
successfully used to generate stationary matter-wave solitons deep
in the spectral gap. Applying the~same technique to~solitons with
different amplitudes can~potentially enable generation
and~steering of~a~mobile high-density soliton. Such a~process of
gap-soliton manipulation may eventually lead to new
possibilities for the preparation of sources for atomic interferometry
\cite{gap_control}.

\section*{Acknowledgement}

This work was partially supported by the Australian Research
Council (ARC). The Australian Centre for Quantum-Atom Optics 
is an ARC Centre of Excellence.

\end{document}